\documentclass[conference]{IEEEtran}
\usepackage{color}
\usepackage{siunitx}
\DeclareSIUnit\gauss{G}
\DeclareSIUnit\oersted{Oe}

\usepackage{hyperref}

\ifCLASSINFOpdf
  \usepackage[pdftex]{graphicx}
\else
\fi
\usepackage{amsmath}
\usepackage{amssymb}

\usepackage{url}

\begin{document}
%
\title{Micromagnetic Simulations for Coercivity Improvement through Nano-Structuring of Rare-Earth Free L1$_0$-FeNi Magnets}

\author{
\IEEEauthorblockN{
Alexander Kovacs\IEEEauthorrefmark{1},
Johann Fischbacher\IEEEauthorrefmark{1},
Harald Oezelt\IEEEauthorrefmark{1},
Thomas Schrefl\IEEEauthorrefmark{1},
Andreas Kaidatzis\IEEEauthorrefmark{2},\\
Ruslan Salikhov\IEEEauthorrefmark{3},
Michael Farle\IEEEauthorrefmark{3},
George Giannopoulos\IEEEauthorrefmark{2}, and
Dimitris Niarchos\IEEEauthorrefmark{2}
}
\IEEEauthorblockA{\IEEEauthorrefmark{1}Center for Integrated Sensor Systems,
Danube University Krems, Wr. Neustadt, AT}
\IEEEauthorblockA{\IEEEauthorrefmark{2}Institute of Nanoscience and Nanotechology, NCRS Demokritos, Athens, GR}
\IEEEauthorblockA{\IEEEauthorrefmark{3}Faculty of Physics and Center for Nanointegration (CENIDE), University of Duisburg-Essen, Duisburg, DE}
}

%


\maketitle

\begin{abstract}
In this work we investigate the potential of tetragonal L1$_0$ ordered FeNi as candidate phase for rare earth free permanent magnets taking into account anisotropy values from recently synthesized, partially ordered FeNi thin films. In particular, we estimate the maximum energy product ($BH$)$_\mathrm{max}$ of L1$_0$-FeNi nanostructures using micromagnetic simulations. The maximum energy product is limited due to the small coercive field of partially ordered L1$_0$-FeNi. Nano-structured magnets consisting of 128 equi-axed, platelet-like and columnar-shaped grains show a theoretical maximum energy product of \SI{228}{\kilo\joule\per\cubic\meter}, \SI{208}{\kilo\joule\per\cubic\meter}, \SI{252}{\kilo\joule\per\cubic\meter}, respectively.
\end{abstract}


%
\IEEEpeerreviewmaketitle

\section{Introduction}
\let\thefootnote\relax\footnotetext{$^\mathrm{\copyright}$~2017 IEEE. Personal use of this material is permitted. Permission from IEEE must be obtained for all other uses, including reprinting/republishing this material for advertising or promotional purposes, collecting new collected works for resale or redistribution to servers or lists, or reuse of any copyrighted component of this work in other works. Published in: \href{https://doi.org/10.1109/TMAG.2017.2701418}{DOI 10.1109/TMAG.2017.2701418}, IEEE
Transactions on Magnetics}
Permanent magnets are found in a large variety of applications~\cite{Lewis2013}. Embedded in consumer electronic devices, computers, speakers/headphones or medical products their demand is steadily increasing. High performance permanent magnets are crucial for operating many green technologies like the motors in electric cars or generators in wind mills. Due to the growing demand and the scarcity of economically suitable rare earth deposits it becomes more important to use sustainable resources for permanent magnets.
In their field of application such magnets are required to govern a large residual magnetization (remanence $B_\mathrm{r}$) and a large resistance to demagnetization (coercivity $H_\mathrm{c}$). Operating temperatures in hybrid cars are typically around \SI{450}{\kelvin} which means that the above mentioned requirements have to be fulfilled at temperatures significantly above room temperature too. The performance of a permanent magnet is expressed through its maximum amount of magnetic energy stored. It is the magnet's optimal product of remanence and coercivity in the second quadrant of its $B$-$H$ hysteresis loop. Reducing or even replacing rare-earth content while keeping this so-called maximum energy product ($BH$)$_\mathrm{max}$ high, ideally even at elevated temperatures, is the major goal.

A commonly used rare-earth magnet is Nd$_2$Fe$_{14}$B which has a theoretical ($BH$)$_\mathrm{max}$ of \SI{\sim 510}{\kilo\joule\per\cubic\meter} (\SI{\sim 64}{\mega\gauss\oersted})~\cite{Skomski2016}. The rare-earth free FeNi with a tetragonal L1$_0$ structure shows a large saturation magnetization of $\mu_\mathrm{0}M_\mathrm{s}=$~\SI{1.5}{\tesla} which translates to a theoretically possible energy product of ($BH$)$_\mathrm{max}=\mu_\mathrm{0}{M_\mathrm{s}}^{2}/4=$~\SI{448}{\kilo\joule\per\cubic\meter} (\SI{\sim 56}{\mega\gauss\oersted}). Here $\mu_0$ is the permeability of vacuum, $\mu_0=\SI{4\pi e-7}{\tesla\meter\per\ampere}$. This makes this particular FeNi phase a potential candidate for permanent magnets, which, in terms of performance, would fill the large gap between the semi-hard AlNiCo and the hard NdFeB magnets. It has to be noted though, that major challenges arise when it comes to engineering such magnets. There is still no cost-efficient method to produce L1$_0$-FeNi in bulk quantities.

This work investigates the potential of L1$_0$-type FeNi via micromagnetic simulations. The following manuscript covers various magneto-crystalline anisotropy constants for L1$_0$-FeNi found in literature. The measured anisotropy constants strongly depend on the used manufacturing techniques. Kojima et al.~\cite{Kojima2012} showed that the magneto-crystalline anisotropy of L1$_0$-FeNi is a function of the chemical ordering.

Miura et. al~\cite{Miura2013} investigated the origin of magneto-crystalline anisotropy in L1$_0$-ordered FeNi alloys with first-principle density-functional calculations. Their calculations reveal a magneto-crystalline anisotropy of $K_\mathrm{u}=\SI{0.56}{\mega\joule\per\cubic\meter}$ within a undisturbed lattice which can be further improved by reducing the in-plane lattice parameter. With their work they provided important guidelines for the choice of buffer layers in thin film L1$_0$-FeNi.

Lewis and co-workers~\cite{Lewis2014,Poirier2015} studied the crystal lattice, microstructure and magnetic properties of the meteorite NWA 6259. Its L1$_0$-FeNi phase is highly ordered and therefore  is regarded as a possible candidate for the application in permanent magnets. They estimated the magneto-crystalline anisotropy of the meteorite to be $K_\mathrm{u}=$~\SI{0.84}{\mega\joule\per\cubic\meter}.

Edstr\"om et al.~\cite{Edstrom2014} investigated the electronic structure and magnetic properties of L1$_0$-FeNi via two different density functional theory approaches and estimated the Curie temperature via Monte Carlo simulations. In their work FeNi showed a magneto-crystalline anisotropy in the range of \SIrange{0.48}{0.77}{\mega\joule\per\cubic\meter}.

Paulev\'e et al.~\cite{Pauleve1968} successfully synthesized chemically ordered L1$_0$-FeNi from the chemically disordered fcc parent phase by fast neutron flux bombardment in an applied magnetic field. N\'{e}el and co-workers~\cite{Neel1964} found that the L1$_0$ phase is forming due to a large concentration of vacancies created by neutron bombardment. With this method they achieved a highly ordered structure with a magneto-crystalline anisotropy of \SI{1.3}{\mega\joule\per\cubic\meter} which is the largest value found in literature.

Kojima and co-workers~\cite{Kojima2012,Kojima2014} successfully fabricated L1$_0$-FeNi films on a CuNi substrate and investigated $K_\mathrm{u}$ and chemical ordering at various substrate temperatures. They showed that increased substrate temperatures lead to higher magneto-crystalline anisotropy (around \SI{0.7}{\mega\joule\per\cubic\meter}). The easy magnetization axis was still in-plane due to the large shape anisotropy of the structure.

Previously, we used a combinatorial sputtering approach to optimize the fabrication of L1$_0$-FeNi films by exploiting a strain-mediated process induced by a suitable buffer-layer. A homogeneous tetragonal Cu$_3$Au seed layer grown on Si(100) wafers and a compositional variation of Cu-Au-Ni buffer layer was used in order to match the lattice constants with the epitaxially grown L1$_0$-FeNi top layer.  Atomic force microscopy showed grains with an average dimension of \SI{380x380x40}{\nano\meter}. The magneto-crystalline anisotropy was measured to be \SI{0.35}{\mega\joule\per\cubic\meter} by ferromagnetic resonance. The above mentioned investigations are taken as basis for our simulations.

\section{Coercivity limited energy product}
Reported anisotropy values for L1$_0$-FeNi both from experiments and simulations suggest that the energy product is coercivity limited (which means $H_\mathrm{c}<M_\mathrm{s}/2$) and the theoretical maximum energy product cannot be reached.

Magnetic reversal in sufficiently large particles without defects happens non-uniformly. The coercive field is then given by~\cite{skomski2008simple}
\begin{equation}
H_\mathrm{c}=H_\mathrm{ani}+H_\mathrm{demag}=2K_\mathrm{u}/J_\mathrm{s}-DJ_\mathrm{s}/\mu_0,
\label{eq:hc}
\end{equation}
where $H_\mathrm{ani}$ is the anisotropy field, $H_\mathrm{demag}$ is the demagnetizing field, $D$ is the demagnetizing factor in the relevant direction~\cite{Aharoni1998} and $J_\mathrm{s}$ the saturation polarization. $D$ is a number ranging from \numrange{0}{1} which represents the shape of a magnet in one particular direction. For a spherical and cubical magnetic body $D_\mathrm{sphere}=D_\mathrm{cube}=1/3$, for a thin film the demagnetizing factors can be approximated~\cite{magpardemag} as $D_\mathrm{film}=0.8$ (\SI{380x380x40}{\nano\meter}) and for an elongated grain as $D_\mathrm{rod}=0.05$ (\SI{40x40x380}{\nano\meter}). A negative contribution of the demagnetizing field $H_\mathrm{demag}$ of a magnetic body has to be compensated by its anisotropy field $H_\mathrm{ani}$ in order to call it a "permanent" magnet. Regarding thin films the demagnetizing factor reaches almost \num{1} in its out-of-plane axis and gives the largest negative contribution to its coercive field.

Hubert et. al~\cite{hubert2008magnetic} introduced the term quality factor $Q$ which is related to the magnetic hardness parameter $\kappa$~\cite{Skomski1999,Coey2010,Coey2012,Hirosawa2015} and represents the ratio between the two opposing energy densities from magneto-crystalline anisotropy $K_\mathrm{u}$ and shape anisotropy $K_\mathrm{d}$:
\begin{equation}
Q=\frac{K_\mathrm{u}}{K_\mathrm{d}},
\label{eq:q}
\end{equation}
where
\begin{equation}
K_\mathrm{d}=\frac{1}{2}\frac{J_\mathrm{s}^2}{\mu_0}.
\label{eq:kd1}
\end{equation}
This reflects the magnetostatic energy density for an infinite thin film magnet. So the quality factor in (\ref{eq:q}) already assumes the worst possible shape for a permanent magnet. If the magnet has a symmetry axis perpendicular to its easy axis we can reformulate (\ref{eq:kd1}) by introducing the shape factors for all directions
\begin{equation}
K_\mathrm{d}=\left(D_{\parallel}-D_{\perp}\right)\frac{1}{2}\frac{J_\mathrm{s}^2}{\mu_0},
\label{eq:kd2}
\end{equation}
where $D_{\parallel}$ and $D_{\perp}$ are the magnet's shape factors parallel and perpendicular to its easy axis, respectively. By plugging in $D_{\parallel}=1-2D_{\perp}$ in (\ref{eq:kd2}), we get
\begin{equation}
K_\mathrm{d}=\left(1-3D_{\perp}\right)\frac{1}{2}\frac{J_\mathrm{s}^2}{\mu_0}.
\label{eq:kd3}
\end{equation}

Here we can see that a shape factor $D_\perp=1/3$ (cube) results in a vanishing demagnetization energy density $K_\mathrm{d}$, which would furthermore translate into a high quality factor $Q$. A perpendicular demagnetization factor $D_\perp<1/3$ on the other hand results in a positive contribution for the coercivity. If $D_\perp\rightarrow 0$ (thin film) we get equation (\ref{eq:kd1}) as pointed out previously. It should be noted that the shape induced anisotropy strictly holds only in particles that reverse uniformly.

Translating the coercivity requirement for reaching the theoretical maximum energy product $H_\mathrm{c}>M_\mathrm{s}/2$ in terms of the quality factor leads to $Q>1/2$. But this requirement is not strict enough for permanent magnets. The logical definition of a permanent magnet implies that it has to act as an actual source of a magnetic field within the absence of an externally applied field. Therefore coercivity has to be greater than the full saturation magnetization $H_\mathrm{c}>M_\mathrm{s}$. This requirement can be translated into $Q>1$. 
Skomski and Coey~\cite{Skomski2016} concluded in their recent work, regarding the required magnetic anisotropy for permanent magnet applications, that the energy product of a magnet is clearly a shape-dependent property and not a material dependent one. A promising candidate material needs a significantly larger magneto-crystalline anisotropy than its squared saturation magnetization $K_\mathrm{u}\gg\mu_0M_\mathrm{s}^2$, assuming its shape is a thin film.

\begin{table}[!t]
\renewcommand{\arraystretch}{1.2}
\caption{Magneto-crystalline anisotropy constant, saturation polarization and the corresponding anisotropy field, theoretical demagnetization field for a cube $\left[\square\right]$ a thin film $\left[-\right]$ and an elongated rod $\left[\mid\right]$. Additionally the quality factor at room temperature is included.}
\label{table_demag}
\centering
\begin{tabular}{|c|c|c|c|c|c|c|c|c|}
\hline
$K_\mathrm{u}$ & $J_\mathrm{s}$ & $\mu_0H_\mathrm{ani}$ & \multicolumn{3}{c|}{$\mu_0H_\mathrm{demag}$ acc. eq.(\ref{eq:hc})} & $Q$ &\\
(\si{\mega\joule\per\cubic\meter}) & (T) & (T) & \multicolumn{3}{c|}{(T)} & (1) & ref.\\
\hline
\hline
\multicolumn{3}{|l|}{L1$_0$-FeNi} & $\left[\square\right]$ & $\left[-\right]$ & $\left[\mid\right]$ & \multicolumn{2}{c|}{} \\
\hline
0.09 & 1.50 & 0.12 & -0.50 & -1.20 & -0.08 & 0.10 & - \\ 
0.35 & 1.50 &  0.47 & -0.50 & -1.20 & -0.08 & 0.39 & - \\ 
0.48 &  1.67 & 0.57 & -0.56 & -1.34 & -0.08 & 0.43 & \cite{Edstrom2014} \\ 
0.54 & 1.30 &  0.83 & -0.43 & -1.04 & -0.07 & 0.80 & \cite{Fukami2013} \\ 
0.70 & 1.50 &  0.93 & -0.50 & -1.20 & -0.08 & 0.78 & \cite{Kojima2012} \\ 
0.77 & 1.72 &  0.90 & -0.57 & -1.38 & -0.09 & 0.65 & \cite{Edstrom2014} \\ 
0.84 & 1.44 &  1.17 & -0.48 & -1.15 & -0.07 &  1.02 & \cite{Poirier2015} \\ 
0.99 & 1.27 &  1.56 & -0.42 & -1.02 & -0.06 &  1.54 & \cite{Edstrom2014} \\ 
1.30 &  1.60 & 1.63 & -0.53 & -1.28 & -0.08 &  1.28 & \cite{Pauleve1968} \\ 
1.35 &  1.30 & 2.08 & -0.43 & -1.04 & -0.08 & 2.01 & \cite{Edstrom2014} \\ 
\hline
\hline
\multicolumn{3}{|l|}{Nd$_2$Fe$_{14}$B} & $\left[\square\right]$ & $\left[-\right]$ & $\left[\mid\right]$ & \multicolumn{2}{c|}{} \\
\hline
4.90 & 1.61 &  6.09 & -0.54 & -1.29 & -0.08 & 4.75 & \cite{Skomski2016} \\ 
\hline
\end{tabular}
\end{table}

If the magneto-crystalline anisotropy cannot be increased any further in order to improve the energy product ($BH$)$_\mathrm{max}$ the introduction of shape anisotropy~\cite{Dubowik1996} through nano-structuring will do.

In the following we will show how nano-structuring may help to increase the energy density product in partially ordered L1$_0$-FeNi.

\section{Method}
In this work the computation of static hysteresis properties are carried out with a finite element energy minimization code~\cite{Schrefl2007}. In combination with the open source framework OpenCL~\cite{stone2010opencl} we are able to compute on massively parallel hardware like GPUs.

For computation the total Gibb's free energy of a magnetic geometry is discretized into finite elements. The mesh size throughout the magnetic sample is \SI{2}{\nano\meter}. We compute the demagnetizing field $H_\mathrm{demag}$ by solving the Poisson's equation for the magnetic scalar potential in the entire space using an algebraic multi-grid method~\cite{Demidov2012}. To calculate the demagnetization curve we subsequently compute the equilibrium states for a decreasing external field $H_\mathrm{ext}$. 

First we investigate demagnetization curves for a single grain with different magneto-crystalline anisotropy constants for L1$_0$-FeNi found in literature. The average grain dimensions were measured with atomic force microscopy to be approximately \SI{380x380x40}{\nano\meter}. The grains easy axis and the external field is assumed to be out of plane.

As second part of this work we investigate the correlation between grain aspect ratio and coercivity for two differently ordered L1$_0$-FeNi materials (shown in Fig.~\ref{fig_ar_vs_hc}). The optimization toolkit Dakota~\cite{Adams2016user} with the optimization algorithm EGO~\cite{Jones1998} has been used. The optimization software proposes an aspect ratio and a simulation script computes the demagnetization curve accordingly and returns the external field value at which the grains magnetization reaches zero. Maximizing $\left|H_\mathrm{c}\right|$ is the optimization's objective. Due to memory limitations the grain's volume has been reduced by a factor of \num{0.3} for aspect ratio optimization. The external field is tilted by \SI{2}{\degree} with respect to the magnets easy axis.

For our aspect ratio optimization we additionally investigate the influence of a \SI{0.75}{\nano\meter} defect shell enclosing the grain. This defect layer has the same material properties as the core material except for its magneto-crystalline anisotropy which is set to zero.

And finally we compute demagnetization curves of a \SI{300x300x300}{\nano\meter} bulk magnet (shown in Fig.~\ref{fig_all-nano-structures}) and how the magnet's coercivity is affected by changing the shape of the residing nano-structured grains.

\section{Results}
\subsection{Single Grain Model}
Atomic force microscopy of FeNi films deposited on a CuAuNi substrate shows grains with average dimensions of \SI{380x380x40}{\nano\meter}. Therefore we performed micromagnetic simulations to compute demagnetization curves for such a platelet shaped L1$_0$-FeNi grain to investigate its reversal process. 

Different magneto-crystalline anisotropy constants were taken into account. These constants found in literature, were acquired experimentally or by density functional theory (DFT).
TABLE~\ref{table_demag} sums up the considered magneto-crystalline anisotropy constants of rare-earth free L1$_0$-FeNi and compares them with the Nd$_2$Fe$_{14}$B magnet. The table also includes saturation polarization $J_\mathrm{s}$, anisotropy field $H_\mathrm{ani}$ and the demagnetizing fields $H_\mathrm{demag}$ of the following three shapes: $\left[\square\right]$ cube, $\left[-\right]$ thin film and $\left[\mid\right]$ elongated rod. Also the quality factor $Q$ for each material is listed too. The grain's easy axis was set parallel to the $z$-axis. The exchange constant $A_\mathrm{ex}$ is assumed to be \SI{10}{\pico\joule\per\meter} throughout all simulation models. 

\begin{figure}[!t]
\centering
\includegraphics[width=\linewidth]{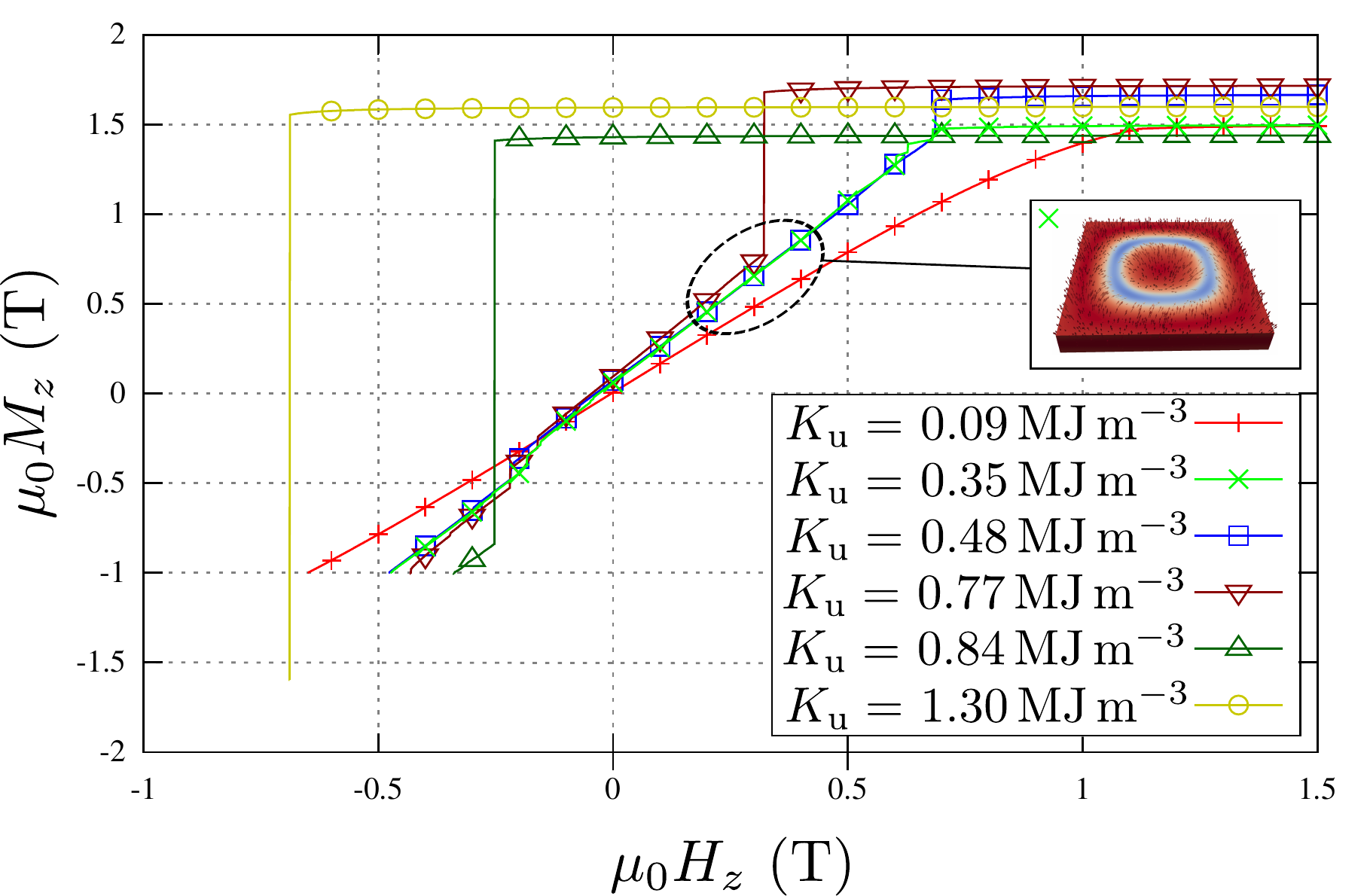}
\caption{(Color Online) Perpendicular magnetization as function of applied field for a single L1$_0$-FeNi grain with dimensions \SI{380}{\nano\meter}$\times$\SI{380}{\nano\meter}$\times$\SI{40}{\nano\meter}. The grain's easy axis points out of plane. Different anisotropy constants found in literature have been considered. Demagnetizing fields start the reversal of the magnetization already in the first quadrant. The inset shows this behaviour for a grain with $K_\mathrm{u}=\SI{0.35}{\mega\joule\per\cubic\meter}$. Regions with magnetization pointing $\uparrow$ are coloured red and $\downarrow$ are coloured blue. Square loops are only achieved for a magneto-crystalline anisotropy constant greater than \SI{0.84}{\mega\joule\per\cubic\meter}.}
\label{fig_1g_kvar}
\end{figure}

Figure~\ref{fig_1g_kvar} gives the magnetization as function of the applied magnetic field for different anisotropy constants. The external field is applied in the same direction as the grains easy axis but with a small tilt of \SI{2}{\degree}. Demagnetizing fields start the reversal of the magnetization already in the first quadrant for materials with an anisotropy lower than \SI{0.84}{\mega\joule\per\cubic\meter}. Single step switching is only achieved for a magneto-crystalline anisotropy constant greater than \SI{0.84}{\mega\joule\per\cubic\meter}. This agrees well with the quality factor condition $Q>1$. Building a permanent magnet out of partially ordered FeNi requires additional shape anisotropy.

\subsection{Aspect Ratio Optimization}
\begin{figure}[!t]
\centering
\includegraphics[width=\linewidth]{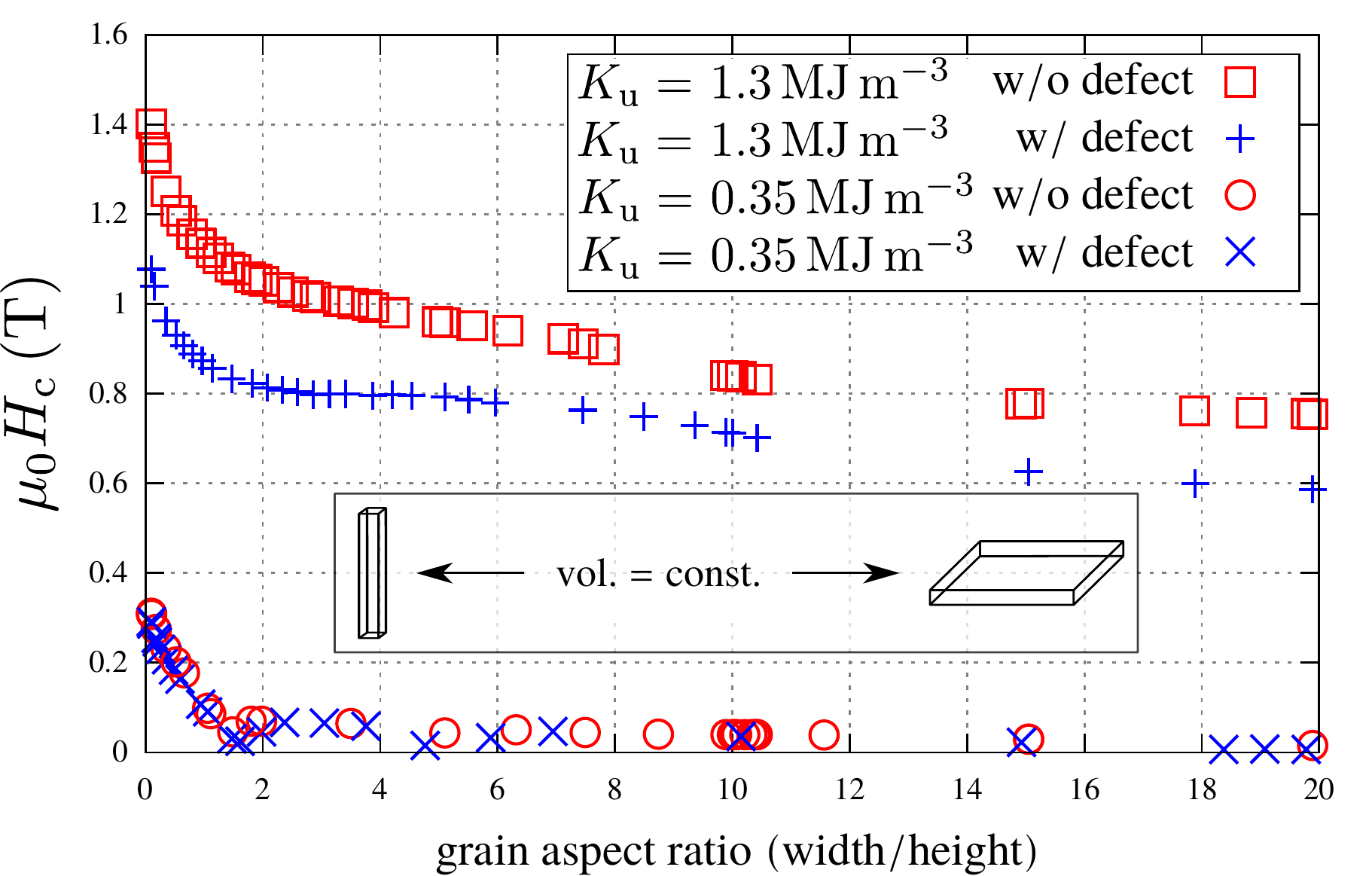}
\caption{(Color Online) Single grain aspect ratio over $\mu_0H_\mathrm{c}$ (field when magnetization reaches zero) with two different anisotropy constants with (blue) and without (red) a \SI{0.75}{\nano\meter} thick isotropic defect shell. As the shape of the grain gets more elongated in easy axis direction (needle shaped) the contribution of shape anisotropy is improving the grain's coercivity.}
\label{fig_ar_vs_hc}
\end{figure}

We additionally investigate the reversal of single grains while changing its aspect ratio (width/height) but keeping the volume constant. In Figure~\ref{fig_ar_vs_hc} we show the computed switching fields for each aspect ratio which has been proposed by the optimization algorithm. The red symbols represent perfect grains (w/o defect) and the blue imperfect grains (w/ defect). We expect that at the surface of the grains the anisotropy is decreasing. Thus we also simulated imperfect grains. The imperfect grains are covered with a \SI{0.75}{\nano\meter} thick layer with zero magneto-crystalline anisotropy. With higher degree of ordering ($K_\mathrm{u} = \SI{1.35}{\mega\joule\per\cubic\meter}$) the decrease in coercivity owing to the defect layer is 20 percent. For low ordering ($K_\mathrm{u} = \SI{0.35}{\mega\joule\per\cubic\meter}$) the relative change in coercivity is much lower (see Figure~\ref{fig_ar_vs_hc}).

All four grain types in Figure~\ref{fig_ar_vs_hc} show increased coercivity at low width$/$height ratios. This is due to the growing shape anisotropy in the direction of the magneto-crystalline easy axis. Adding the defect layer results in a \SI{10}{\percent} reduction of $H_\mathrm{c}$, regardless of both the grain's $K_\mathrm{u}$ and aspect ratio.

\subsection{Bulk Model}
\begin{figure}[!t]
\centering
\includegraphics[width=\linewidth]{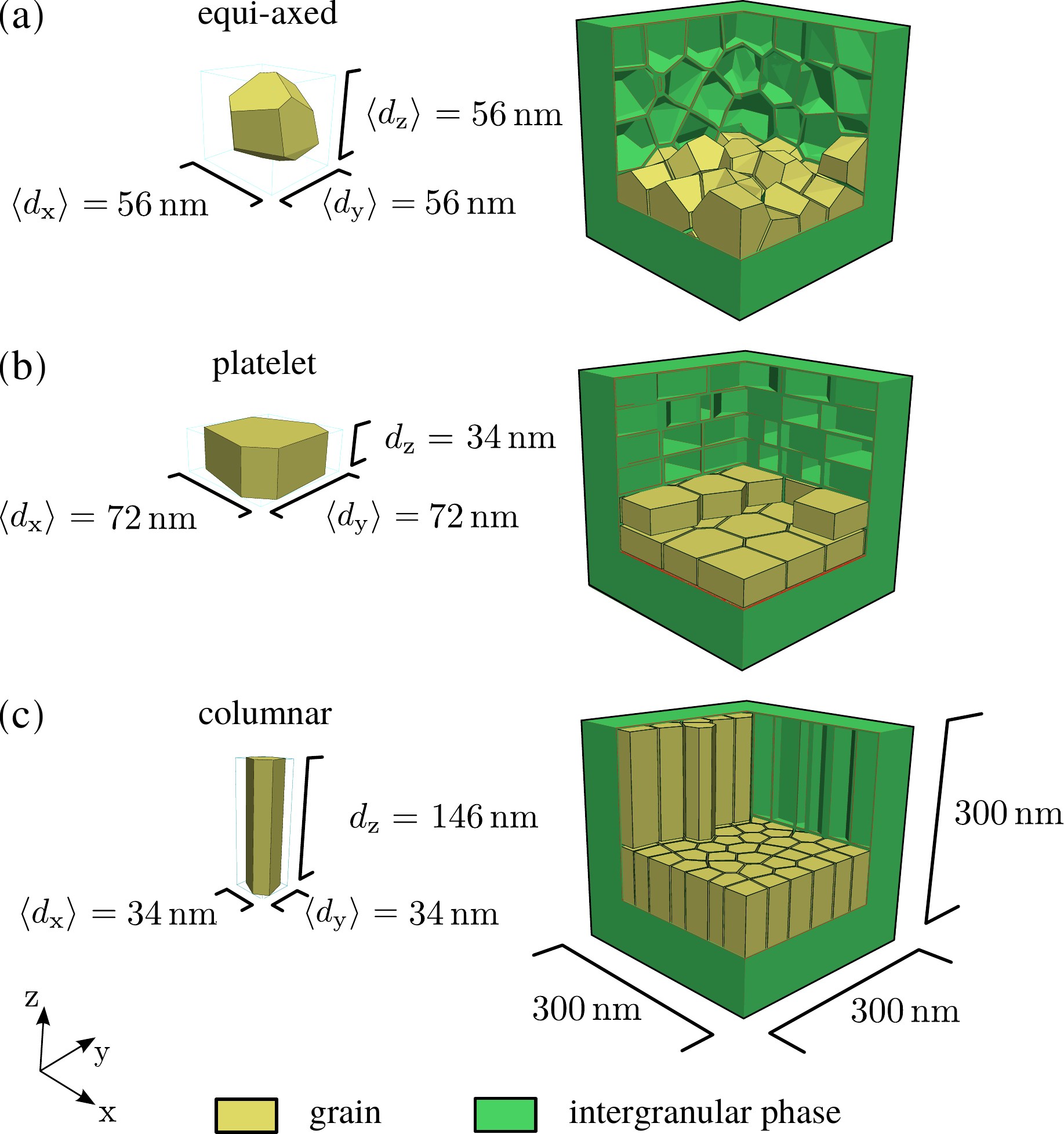}
\caption{(Color Online) Three different nano-structured bulk simulation models (right column) and their included grain shape (left column). Each model consists of \num{128} Voronoi grains: (a) equally grown in each axis-direction (b) quenched platelet and (c) columnar shaped. Each grains easy-axis is set parallel to z-axis. The grains (yellow) are embedded in a non-magnetic intergranular phase (green). Together they are forming a \SI{300x300x300}{\nano\metre} bulk sample. Total magnetic volume stays the same in all three geometries. Dimensions for single grains on the left hand side are average sizes.}
\label{fig_all-nano-structures}
\end{figure}

In order to investigate how the grain shape influences the coercivity in larger L1$_0$-FeNi systems we constructed a \SI{300x300x300}{\nano\meter} model filled with 128 grains. Between the grains a non-magnetic grain boundary phase is taken into account, which occupies \SI{18}{\percent} of the volume. We considered three different shapes for the grains which reflect the average demagnetization factor for a cube, a thin film or an elongated rod (see Fig.~\ref{fig_all-nano-structures}). The multi-grain bulk models were generated with the polycrystalline generator Neper~\cite{Quey2011} with additional geometry adaptations and mesh generation carried out with the computer aided design software Salome~\cite{salome}.

Figure~\ref{fig_all-nano-structures} shows average dimensions of single grains and the corresponding nano-structured bulk models we investigate. The average grain dimensions are \SI{56x56x56}{\nano\meter}, \SI{72x72x34}{\nano\meter}, and \SI{34x34x146}{\nano\meter} for the equi-axed, platelet and columnar grain shape, respectively. Figure~\ref{fig_all-nano-structures}a shows the equi-axed grain model in which each grain has been constructed with 3D Voronoi tesselation. The platelet-shaped grains on the other hand (Fig.~\ref{fig_all-nano-structures}b) are constructed layer by layer where each level of these quenched grains includes a separate 2D Voronoi tessellation. The simulation model of the columnar shaped grains (Fig.~\ref{fig_all-nano-structures}c) consists of one single 2D Voronoi tessellation extruded once and a duplicate of the first layer placed on top of its original seperated by a grain boundary.

\begin{figure}[!t]
\centering
\includegraphics[width=\linewidth]{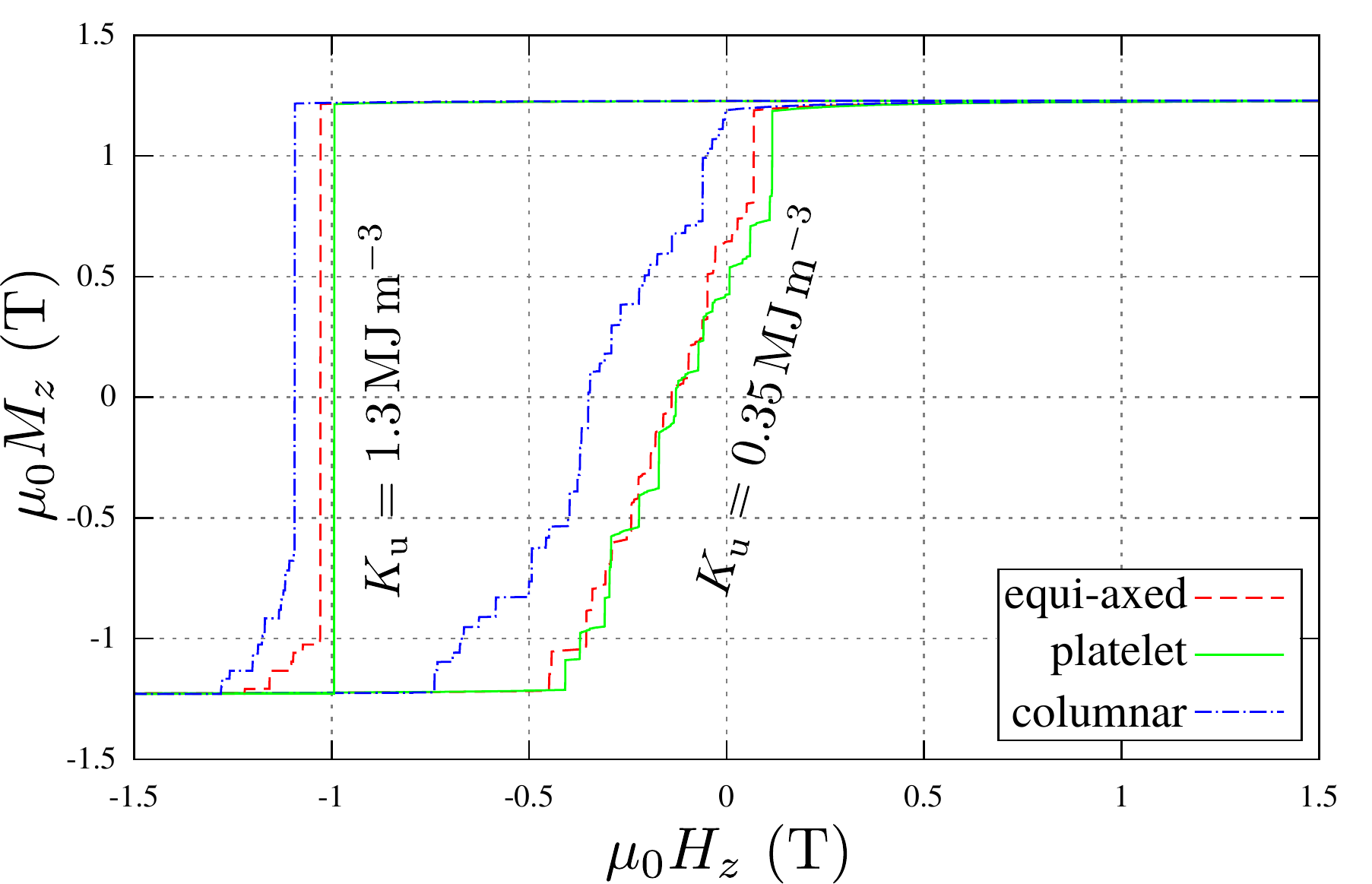}
\caption{(Color Online) Reversal curves of nano-structured magnets with (dashed) equi-axed grains, (solid) platelet grains and (dot-dashed) columnar grains. We assume that the grains easy-axes are perfectly aligned. The three curves on the left hand side show results for L1$_0$-FeNi with an anisotropy constant of $K_\mathrm{u}=\SI{1.3}{\mega\joule\per\cubic\meter}$. The three curves on the right hand side are for $K_\mathrm{u}=\SI{0.35}{\mega\joule\per\cubic\meter}$.}
\label{fig_shape_kall}
\end{figure}

For computing the demagnetization curve we assume the following two different magneto-crystalline anisotropy constants: 1) a moderate anisotropy of \SI{0.35}{\mega\joule\per\cubic\meter} which corresponds to a chemical order parameter of \num{0.25} according to Kojima~\cite{Kojima2012} and 2) an anisotropy of a chemically highly ordered FeNi phase with \SI{1.3}{\mega\joule\per\cubic\meter}~\cite{Pauleve1968}. The computed reversal curves of the above mentioned nano-structured magnets are shown in Figure~\ref{fig_shape_kall}. The external magnetic field $H_\mathrm{ext}$ acts in negative $z$-axis direction but deviates by \SI{2}{\degree} from the $z$-axis. Shape anisotropy increases the coercive field of the columnar model by \SI{10}{\percent} compared to the platelet model for $K_\mathrm{u}=\SI{1.3}{\mega\joule\per\cubic\meter}$ and by \SI{170}{\percent} for $K_\mathrm{u}=\SI{0.35}{\mega\joule\per\cubic\meter}$.

We further compute the energy product of these specific multi-grain structures taking into account the macroscopic demagnetizing factor $D=1/3$ of the cubic sample. For a volume fraction of FeNi of \num{82} percent the theoretical maximum energy density product is approximately \SI{300}{\kilo\joule\per\cubic\meter} (\SI{\sim 38}{\mega\gauss\oersted}). 
The micromagnetic results for $K_\mathrm{u}=\SI{0.35}{\kilo\joule\per\cubic\meter}$ show lower values which indicate that the energy density product is coercivity limited. The computed ($BH$)$_\mathrm{max}$ values are \SI{251}{\kilo\joule\per\cubic\meter} (\SI{\sim 32}{\mega\gauss\oersted}), \SI{225}{\kilo\joule\per\cubic\meter} (\SI{\sim 28}{\mega\gauss\oersted}), and \SI{201}{\kilo\joule\per\cubic\meter} (\SI{\sim 25}{\mega\gauss\oersted}), for needle like, equi-axed, and platelet like grains, respectively.

\section{Conclusion}
The chemically ordered L1$_0$-FeNi phase has a large saturation magnetization with a theoretically high maximum energy product. But reported anisotropy values from simulations and experiments suggest that the energy product is coercivity limited. Performed hysteresis loop simulations of a single \SI{380x380x40}{\nano\meter} grain show that a magneto-crystalline anisotropy of at least \SI{0.84}{\mega\joule\per\cubic\meter} is required to get hard-magnetic behaviour. This is also reflected by the requirement of a quality factor $Q$ greater than one as listed in TABLE~\ref{table_demag}. If the magneto-crystalline anisotropy cannot be increased any more the introduction of shape anisotropy has to be taken into consideration. 
Optimizing the grain aspect ratio at constant grain volume shows that the best shape is the columnar-shaped one.
This shape also performs best regarding the energy product in a \SI{300x300x300}{\nano\metre} bulk model consisting of 128 elongated grains. It has to be mentioned that a method to fabricate such nano-structured magnets in a cost efficient way still has to be invented.

\section*{Acknowledgment}
Work supported by funding from the European Union’s Horizon 2020 NMBP23-2015 research No. 686056 (NOVAMAG).

\bibliographystyle{IEEEtran}

\bibliography{IEEEabrv,BIBLIO}

\end{document}